\documentclass[10pt,a4paper,twoside]{article}
\usepackage{epsfig}
\usepackage{baltlat6}
\usepackage{array}
\usepackage{here}
\usepackage{multirow}
\usepackage{graphicx}

\begin{document}
\ \ \vspace{0.5mm} \setcounter{page}{53}

\titlehead{Baltic Astronomy, vol.\,22, 53-65, 2013}

\titleb{INVESTIGATION OF THE EARTH IONOSPHERE \\
USING THE RADIO EMISSION OF PULSARS\\ }

\begin{authorl}
\authorb{O.M. Ulyanov}{},
\authorb{A.I. Shevtsova}{},
\authorb{D.V. Mukha}{},
\authorb{A.A. Seredkina}{}
\end{authorl}
\begin{addressl}
\addressb{}{Department of Astrophysics, Institute of Radio Astronomy of  NAS of Ukraine,
\\  Krasnoznamennaya str. 4, Kharkov 61002, Ukraine;
oulyanov@rian.kharkov.ua}
\end{addressl}

\submitb{Received: 2012 November 13; accepted: 2012 November 28}

\begin{summary}
The investigation of the Earth ionosphere both in a quiet and a
disturbed states is still desirable. Despite recent progress in
its modeling and in estimating the electron concentration along
the line of sight by GPS signals, the impact of the disturbed
ionosphere and magnetic field on the wave propagation still
remains not sufficiently understood. This is due to lack of
information on the polarization of GPS signals, and due to poorly
conditioned models of the ionosphere at high altitudes and strong
perturbations. In this article we consider a possibility of using
the data of pulsar radio emission, along with the traditional GPS
system data, for the vertical and oblique sounding of the
ionosphere. This approach also allows to monitor parameters of the
propagation medium, such as the dispersion measure and the
rotation measure using changes of the polarization between pulses.
By using a selected pulsar constellation it is possible to
increase the number of directions in which parameters of the
ionosphere and the magnetic field can be estimated.

\end{summary}


\resthead{Investigations of the Earth's Ionosphere using Pulsar
Radio Emission} {O.M. Ulyanov, A.I. Shevtsova, D.V. Mukha, A.A.
Seredkina}

\sectionb{1}{INTRODUCTION}

The observable radio pulses are generated in the space surrounding
a neutron star, which is usually called the pulsar magnetosphere.
These radio pulses have a number of specific features, but for
using them as sounds of the propagation medium it is essential for
them to possess a broadband spectrum, strictly periodic pulse
signals and a polarized radio emission. The spectrum of radio
pulses of pulsars extends from the decameter to the millimeter
ranges, their periods are from milliseconds to seconds, and the
degree of linear polarization of the strongest pulses is close to
100\%. These three factors make it possible to use the strongest
pulses to sound the propagation medium, including the Earth
ionosphere.

The aim of this work is to use a polarized periodic pulsar
radiation for sounding the Earth ionosphere. In this case, the
pulsar emission will be used as an additional source for sounding
the ionosphere to monitor rapid fluctuations of the Earth magnetic
field along the line of sight.

To solve this problem in a general case of elliptically polarized
radio emission propagating along the line of sight, it is
advisable to use the Faraday effect which causes rotation of the
polarization plane of this emission in the presence of a magnetic
field component parallel to the line of sight. It will be shown
that the evaluation of rapid fluctuations in the magnetic field
near the Earth will require the estimates of the rotation measure
(hereafter $RM$) for every single pulse of the pulsar. Then, by
checking this parameter and by using a model of the propagation
medium, it is possible to obtain estimates of rapid fluctuations
of the magnetic field near the Earth, and to find possible
deviations of the ionospheric parameters from the average.

\sectionb{2}{THE MODEL CONCEPTION OF THE POLARIZED PULSAR RADIO \\
EMISSION}

Despite more than a forty-year investigation of the pulsar radio
emission (PRE) (Hewish et al. 1968, Pilkington et al. 1968), the
mechanism of their coherent radio emission remains unclear
(Melrose \& Luo 2008). Observational data in different frequency
ranges show that radio emission of pulsars is polarized (McKinnon
2009). Typically, the linear polarization dominates (Suleimanova
\& Pugachev 2002), however, the component with circular
polarization is also involved (Melrose 2003).

We have simulated radio pulses with the repetition period
corresponding to the rotation period of a pulsar, $P =$ 1 s. In
order to simulate the most common case considering the presence of
both linear and circular polarizations in PRE, we have included
the elliptical polarization into the model. The model signal was
generated as follows:

(1) First, using a random number generator that provides a uniform
distribution of random variables, we generated the carrier
frequencies $f(k)$ which were then uniformly distributed in
bandwidths. As the central frequency, we accepted the decametric
frequencies at 20 MHz and 30 MHz. These frequencies are commonly
used in observations on the UTR-2 radio telescope (Ul'yanov et al.
2006, 2008, 2012);

(2) A random number generator has formed the first array of
amplitudes $A_0(k)$ which had the Rayleigh distribution with the
given r.m.s. deviation ($\sigma_1$). Each amplitude in the array
matched its radio frequency, see Equation (1) below;

(3) The second amplitude array $B_0(k)$, which had a similar
distribution but different standard deviation ($\sigma_2$), was
consistent with the orthogonal (relative to the first array)
linearly polarized component of the radiation on the same carrier
frequency as in the item (2). The ratio of standard deviations
($\sigma_1/\sigma_2$), taken for the two Rayleigh amplitude
distribution laws determined the compression ratio of the ellipse.
Commonly, this ratio is considered to be not higher than 0.5;

(4) Similarly, a white noise was generated with a normal
distribution and added to the signal. For the greater clarity and
more accurate interpretation of the results, we used the
signal/noise ratio $(S/N)>$ 20;

(5) The amplitude of the set of noise-like carrier frequencies was
modulated. The envelope $G(t)$ of the modulated signal had the
Gaussian shape with width not exceeding 10\% of the pulsar period.
Thus, a pulse nature of the pulsar signal was simulated in the
observer frame, see Equation (2) below;

(6) We specified a variation of the position angle (PA) along the
average profile of the pulse envelope, using a function which is
smooth and slowly varying along the pulse profile $\chi(t)=\arctan
\left({(t-t_{center})}/ {m} \right)$. The denominator in the
argument of this function could be changed, simulating different
slope of the PA trend in the pulse window. A similar shape of the
PA trend is actually observed in the polarized radio emission of
pulsars in different frequency ranges (Karastergiou and Johnston
2006);

(7) Due to limited resources of the computer, in both ranges (20
MHz and 30 MHz), the bandwidths were chosen to be relatively small
and equal to 4.8 kHz;

(8) To simplify the solution of the direct and inverse problems,
we used the analytical form for the representation of model
signals (Marple 1999).

Thus, two independent channels (A and B) were formed in which the
simulated pulse signals had an orthogonal linear polarization and
PA of a given trend. The resulting signal possessed a given
ellipticity coefficient and a specified $(S/N)$ ratio.

These model concepts of the signal are formed on the assumption
that the propagation channels have relatively narrow bands. In
addition, we believe that the tested signals have a fixed
ellipticity coefficient on the altitudes of the pulsar
magnetosphere which are higher from the surface than the critical
polarization radius (Petrova 2006a,b). We should also consider the
effect of retardation, the effect of aberration, the dispersion
delay of signals in a cold plasma, and the Faraday effect in
propagation at the altitudes exceeding the critical radius of
polarization (Wang et al. 2011). The first two effects are the
kinematic ones and, as it was shown by Ulyanov (1990), at the
first approximation they compensate each other. The Faraday effect
is related to the rotation of the polarization plane along the
line of sight. It is caused by different phase propagation
velocity of the ordinary ($O$) and extraordinary ($X$) waves
(Ginzburg 1967) which arise due to different refractive indices of
these waves in a cold anisotropic plasma.

The following equations represent the signals themselves. For the
index $k \leq N$, where $f(N)$ is the Nyquist frequency, analytic
signal is:
\begin{equation}                                                          
\begin{array}{l}
 {\dot E}_{0x} (k,t)=A_{0} (k) \cdot e^{-i(2\pi f(k) \tau(t))} \\
 {\dot E}_{0y} (k,t)=B_{0} (k) \cdot e^{-i(2\pi f(k) \tau(t)- \pi / 2)}
\end{array},
\end{equation}
where $A_{0} (k)=\sqrt {-2 \sigma _{1} \ln(1-x (k))}$ and $B_{0}
(k)=\sqrt {-2 \sigma _{2} \ln(1-y (k))}$ are the sets of
amplitudes for each channel respectively, $x(k)$ and $y(k)$ are
the independent random values with uniform distribution from 0 up
to 1, $f(k)$ is the set of frequencies, $\tau (t) $ is the time
sample. For $k
> N; ~~~{\dot E}_{0x,0y} ({k,t})=0$. Then the signal is summed to
the index $k$, and in the form of an impulse to the presence of
noise is written as:
\begin{equation}                                                          
\begin{array}{l}
{\dot E}_x(t)= G(t) \cdot \sum \limits_{k} {\dot E}_{0x} (k,t) +
N_x(t)\\ {\dot E}_y(t)= G(t) \cdot {\sum \limits_{k}} {\dot
E}_{0y} (k,t) + N_y(t)
\end{array},
\end{equation}
where $G(t)$ is the Gaussian envelope function, $N_{x,y}(t)$ is
the white noise.

\sectionb{3}{THE MODEL CONCEPTION OF THE PROPAGATION MEDIUM}

At this stage of study, the best solution is to present the
propagation medium in a form of layers with changable parameters
in accordance with the known analytical laws. In this approach,
each layer in the propagation medium is convenient to represent by
an eikonal equation $\nabla \varphi(\omega) = n(\omega) \vec k
(\omega)$, where $\varphi (\omega)$ is the signal phase, $n
(\omega)$ is the refractive index of the medium, $\vec k (\omega)$
is the wave vector. This equation is applicable to the situation
when the propagation medium still has no non-linear effects. By
using Taylor series expansion of the refractive index $n
(\omega)$, in the presence of the longitudinal component of the
magnetic field along the line of sight, it is possible to keep
only the first three terms which consider the influence of the
propagation medium. These terms are additive.

The key moment when using an eikonal equation for modeling the
propagation medium is to present the refractive indices of the
ordinary and extraordinary waves (Ginzburg 1967). With the
quasi-longitudinal propagation, according to Zheleznyakov (1977,
1997), we define the ordinary wave as a wave which has the right
circular polarization while the direction of its wave vector
coincides with the direction of the magnetic field vector, and the
magnetic field itself is directed to the observer. Accordingly,
the extraordinary wave will have the left circular polarization
under the same conditions. If the magnetic field is directed from
the observer to the source and the wave vector has the opposite
direction, the ordinary wave will have the left circular
polarization, and the extraordinary wave - the right circular
polarization. Now we define the refractive indices in the
propagation medium (one of which, $n_O(\omega)$, corresponds to
the ordinary wave and another one, $n_X(\omega)$, to the
extraordinary wave) by the following equation: ${n_{O,X} (\omega)}
=\sqrt{1-{\omega_p^2}/ {\omega(\omega \mp \omega_H)}}$.

Let us consider under which conditions and/or limitations it is
fair to use the equations of a quasi-longitudinal propagation. The
determining value of the analysis of radio wave modes in the
presence of magnetic field is a characteristic $u=(\omega_H/
\omega)^2= (e |\vec{B }|/(m_e c~ \omega ))^2$. For the propagation
conditions in the galactic interstellar plasma at the lowest
observation frequency of 20 MHz, where the average values of the
magnetic induction $ \langle | { {{\vec B}_{ISM} }|}\rangle  ~
\approx ~ 1 ~\mu G$, the parameter u is equal to $u = 1.96\cdot
10^{-14}$. Thus it is evident that for all angles between the wave
vector $\vec {k}$ and ${\vec {B}}_{ISM}$, except of the angles
$\angle \vec {k} {\vec B}_{ISM}$ close to $\pm \pi/2$, the
propagation of waves at frequencies above 20 MHz has a
quasi-longitudinal nature with circular normal modes.

Propagation conditions in the Earth ionosphere should be
considered in more detail. To determine the ellipticity of normal
modes of the radio emission in a cold anisotropic plasma, we use
the following equation (Zheleznyakov 1997):

\begin{equation}
K_{O,X}~=~ \frac{2 \sqrt{u} \times \cos(\angle \vec{k}               
{\vec B}_I) } {u \times {\sin^2(\angle \vec{k} {\vec B}_I)}~\pm~
\sqrt{u^2 \times \sin^4({\angle \vec{k} {\vec B}_I})~+~ 4 u \times
\cos^2({\angle \vec{k} {\vec B}_I})}},
\end{equation}
where ${\vec B}_I$ is the magnetic induction vector in the Earth
ionosphere.

The magnitude and direction of $\vec{B_I}$ which is required for
the analysis, depend on the geographical coordinates of the phase
center of the receiving radio telescope. All the data from
Equation (3) will be given for a mid-latitude radio telescope
UTR-2 with the following coordinates: $49^\circ~ 38' ~17.6''$ N,
$36^\circ~ 56'~ 28.7''$ E, and the altitude 180 m.On the date of
observation, the magnetic induction vector in the phase center of
UTR-2 had the following parameters $ |{\vec{B}_I}|~=~50419 \times
10^{-9}$ T, i.e., about 0.5 G. The vertical and horizontal
components had the values $|\vec{B}_{Iv}|~=~46311.6 \times
10^{-9}$ T and $ |{\vec{B}_{Ih}}|~=~19932.7 \times 10^{-9}$ T,
respectively. In this case, the horizontal component of the vector
corresponds to the South - North direction with a slight deviation
(about $7^\circ$) to the East. Thus, the inclination angle of the
magnetic induction is about  $23^\circ $. Since in the first
approximation the Earth magnetic field has a dipole character, the
value of its magnetic intensity decreases with height as $
{|\vec{B}_I|} \sim 1/R^3$, where $R$ is the distance from the
center of the magnetic dipole (the center of the Earth). Daily
variations of the Earth magnetic field for undisturbed conditions
at middle latitudes do not exceed ${|\delta \vec{B}_I|} \leq 50
\cdot 10^{-9}$ T.

Then, by substituting the above parameters to Equation (3), we
obtain the values of ellipticity coefficients $K_{O}~=~ -1/{K_{X}}
$ for the normal modes of radio emission in the phase center of
the UTR-2 radio telescope at any angle between $\vec {k}$ and
$\vec{B}_I$ (see Figure 1).
\begin{figure}[h] \vbox{
\centerline{\psfig{figure=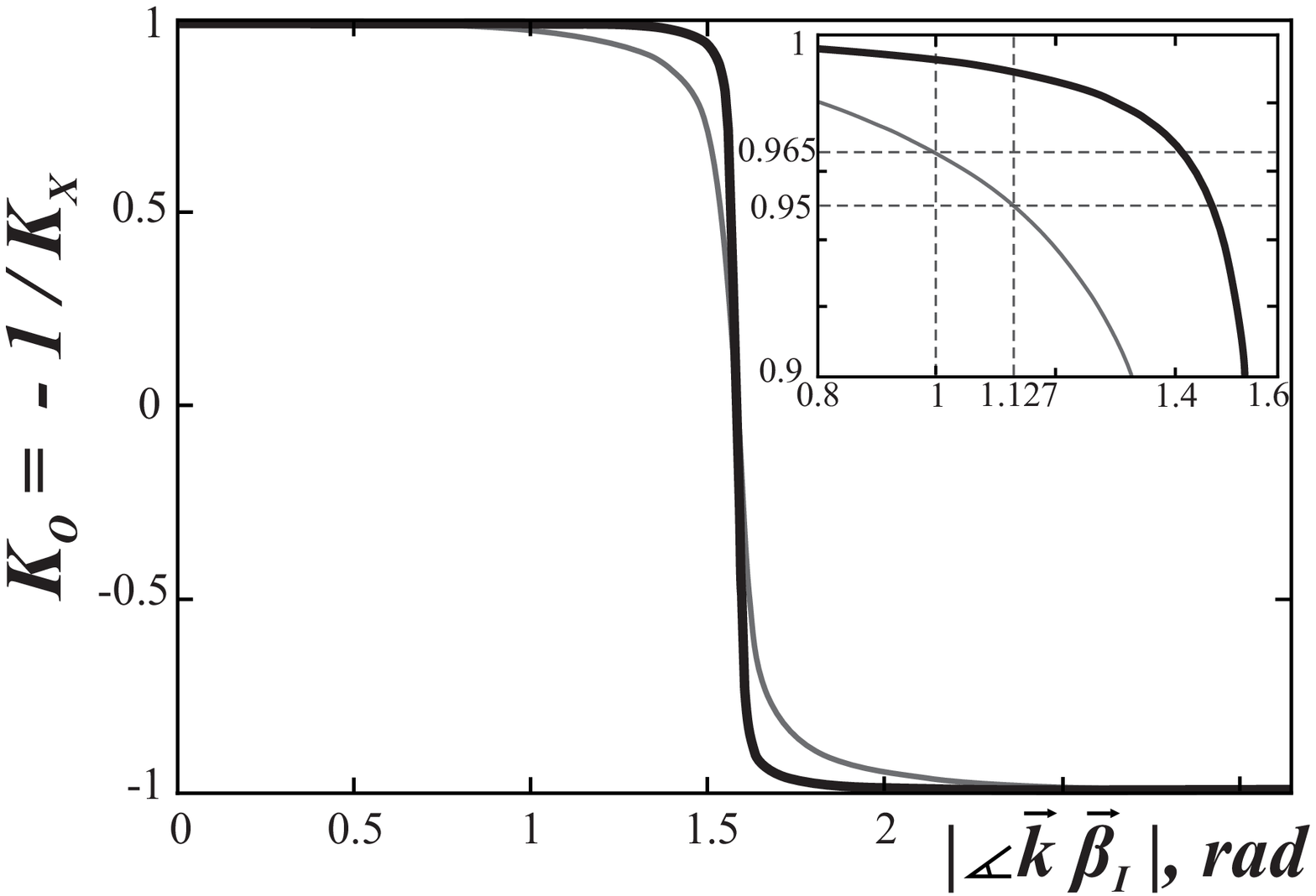,width=60mm,angle=0,clip=}}
\vspace{1mm} \captionb {1}{Ellipticity coefficient for normal
modes in the angular range  $0 \leq \angle |\vec {k} \vec{B}_I
|\leq \pi $ }at the frequencies 23.7 MGz (thin line) and 111 MGz
(thick line).}
\end{figure}

From Figure 1 it is evident that in the angular range $0^{\circ}
\leq ({\angle \vec{k} {\vec B}_{I}}) \leq 65^{\circ} ~ \bigcup ~
\\
 115^{\circ}  \leq ({\angle \vec{k} \vec{B}_{I}}) \leq
180^{\circ}$ both ellipticity coefficients at frequencies $F_{c}
\geq 23.7$ MHz exceed by absolute value the 0.95 level for the
considered conditions. Such values of ellipticity coefficients of
the normal modes correspond to the case of quasi-longitudinal
propagation in a given range of angles at frequencies that exceed
the above value. Therefore, the further analysis will be done from
the point of view of a quasi-longitudinal propagation with the
circular normal modes.

For a transverse wave propagating along the axis $z$, the eikonal
equation is:
\begin{equation}
\frac{d \varphi_{O,X}(\omega)}{dz}~=~n_{O,X}(\omega) k(\omega) ~=~     
n_{O,X}(\omega)\frac{\omega}{c}{}.
\end{equation}

The expansion of Equation (4) in Taylor series gives the phase of
the signal detected in any space-time point on the line of
sight:
\begin{equation} \varphi_{O,X}(\omega) \approx  \frac{\omega
L}{c}  -
\frac{1}{\omega} \frac{2 \pi e^{2}}{m_{e} c } \int_{0}^{L} {N_e        
(z)} dz \mp \frac {1}{\omega^2} \frac {2\pi e^3}{{m_e}^2 c^2}
\int_{0}^{L} N_e(z) |\vec{B}|\cos({\angle \vec k \vec B})  dz{~},
\end{equation}
where $c$ is the speed of light, $e$ is the electron charge, me is
the electron rest mass, $L$ is the distance between a pulsar and
an observer or the layer thickness, $N_e(z)$ is the electron
concentration at the line of sight, $|\vec{B}|\cos({\angle \vec k
\vec B})$ is the magnetic field intensity along the propagation
path.

In Equation (5), the terms are grouped so as to separate in the
propagation in vacuum, the dispersion delay and the Faraday
effect. Equations (6) and (7) show the phase shift caused by these
two effects, respectively:
\begin{equation} \alpha (DM,\omega)
~\approx~   \frac{1}{\omega} \frac{2 \pi e^{2}} {m_{e} c}
\int_{0}^{L} {N_e (z)} dz ~=~\frac{e^2}{m_e c}DM \frac {1}             
{f}{~},
\end{equation}
\begin{equation}
\psi_{O,X}(RM,\omega) ~\approx~ \mp~ \frac{1}{\omega^2} \frac{2\pi
e^3}{m^2 c^2} \int_{0}^{L}{N_e (z)} |\vec{B}|\cos({\angle \vec k \vec B})  dz ~=~ \mp {RM            
\lambda^2}~,
\end{equation}
where $DM =\int_{0}^{L}{N_e (z)} dz$ is the dispersion measure, $f
= \omega /(2\pi)$, $\lambda$ is the wavelength, $RM = \frac
{e^3}{2 \pi m^2 c^4}\int_{0}^{L} {N_e (z)} |\vec{B}|\cos({\angle
\vec k \vec B})dz$ is the rotation measure.

Now we can move on to the matrix description of the layered model
of the propagation medium. The matrices, which are responsible for
the dispersion delay in a cold isotropic plasma and for the
Faraday rotation in the presence of magnetic field commute.
Therefore, the solution of both the direct and the inverse
problems is greatly simplified due to a possibility to enter and
to offset the impact of these two propagation effects separately.
Accordingly, the inverse problem can be solved by using the matrix
inversion, used in the construction of the direct problem.

Herein we give the equations in a matrix form for the signals that
can be detected by using the so-called Wave Form (WF) receivers
(Zakharenko et al. 2007) in a homogeneous layers of the
interstellar medium (ISM) and the Earth ionosphere:
\begin{eqnarray}                                                      
\left[
\begin{array}{c}
  {{\dot E}_{x}}^{space}(\omega) \\
  {{\dot E}_{y}}^{space}(\omega) \\
  {{\dot E}_{z}}^{space}(\omega)
\end{array}
\right]= D \times  {T_{lc}}^{-1} \times R
 \times T_{lc} \times PA \times  \left[
\begin{array}{c}
  {{\dot E}_x}(\omega) \\
  {{\dot E}_y}(\omega) \\
  {{\dot E}_z}(\omega)
\end{array}
\right]{ ~},
\end{eqnarray}
${\dot E}_{x,y,z}^{space}(\omega)$ are the signals in the free
space under ionosphere, ${\dot E}_{x,y}(\omega)$ are created
signals from Eq.(2) ($\dot E_z (\omega)=0$), $T_{lc}$ is the
transfer matrix from linear basis to circular basis,
 ${T_{lc}}^{-1}=T_{cl}$ is the
inverse transfer matrix.
\begin{eqnarray}
T_{lc}= \frac{1}{\sqrt{2}} \left[
\begin{array}{ccc}
  1 & i & 0 \\                                                              
  1 & -i & 0 \\
  0 & 0 & \sqrt{2}
\end{array}\right],~~
T_{lc}^{-1}= \frac{1}{\sqrt{2}} \left[
\begin{array}{ccc}
  1 & 1 & 0 \\
  -i & i & 0 \\
  0 & 0 & \sqrt{2}
\end{array}\right] \nonumber ,
\end{eqnarray}
\begin{eqnarray}                                                            
D~=\left[\begin{array}{ccc}
  e^{-i\alpha(DM,\omega)} & 0 & 0 \\
  0 & e^{-i\alpha(DM,\omega)} & 0 \\
  0 & 0 & 1
\end{array}\right]
, R~=\left[
\begin{array}{ccc}
  e^{i\psi_O(RM,\omega)} & 0 & 0 \\
  0 & e^{i\psi_X(RM,\omega)} & 0 \\
  0 & 0 & 1
\end{array}\right]
\nonumber
\end{eqnarray}
are the matrices of phase shifts caused by the dispersion delay
and the Faraday effect, $ PA=\left[
\begin{array}{ccc}
 \cos(\chi(t)) & \sin(\chi(t)) & 0 \\
  -\sin(\chi(t)) & \cos(\chi(t)) & 0 \\
  0 & 0 & 1
\end{array}\right]$
 is the rotation matrix with simulated $\chi(t)$.

We should also focus on compensation of the influence of the
underlying surface near the interface of two media (air and land)
and the redistribution of the intensity of polarization components
received by a radio telescope, which appears due to the presence
of the underlying surface. So, under ${\dot
E}_{x,y,z}^{space}(\omega)$ we mean the corresponding components
of the field at the output of the ionosphere in the plane of the
radio telescope phase center that is orthogonal to $\vec {k}$ Then
we extend Equation (8) and take into account the actual ground
conductivity by using of the Fresnel reflection coefficients
(Eisenberg et al. 1985; Schelkunoff \& Friis 1952). These
coefficients are different for perpendicular (s) polarization
(sticks out of or into the plane of incidence z0y) and parallel
(p) polarization (lies parallel to the plane of incidence)
components of the electric field: $$\dot\rho_s= \frac{\cos(\angle
z'z)- \sqrt{\dot \varepsilon-{\sin^2(\angle z'z)}}}{ \cos(\angle
z'z)+\sqrt{\dot \varepsilon-{\sin^2(\angle z'z)}}}~,~ \dot \rho_p=
\frac {\dot \varepsilon \cos(\angle z'z)- \sqrt{\dot
\varepsilon-{\sin^2(\angle z'z)}}}{\dot\varepsilon \cos(\angle
z'z)+\sqrt{\dot \varepsilon-{\sin^2(\angle z'z)}}}~.$$

Equation (8), taking into account the underlying surface (we
assume that it is flat), is transformed into the following form:

\begin{eqnarray}                                                      
\left[
\begin{array}{c}
  {{\dot E}_{x'}}^{rec}(\omega) \\
  {{\dot E}_{y'}}^{rec}(\omega) \\
  {{\dot E}_{z'}}^{rec}(\omega)
\end{array}
\right]= COS \times(\|1\|+REF)  \left[
\begin{array}{c}
  {{\dot E}_x^{space}}(\omega) \\
  {{\dot E}_y^{space}}(\omega) \\
  {{\dot E}_z^{space}}(\omega)
\end{array}
\right]{ ~},
\end{eqnarray}
where $x'~;~y'~;~ z'$ are the new coordinate axes in the
laboratory frame, ${\dot E}_{x',y',z'}^{rec}(\omega)$ are the
signals at the receiver, $\parallel 1 \parallel~=~ \left[
\begin{array}{ccc}
  1 & 0 & 0 \\
  0 & 1 & 0 \\
  0 & 0 & 1
\end{array}\right]$ is the unit matrix (transform directed signal matrix),
$ COS=\left[
\begin{array}{ccc}
 \cos(\angle x'x) & \cos(\angle x'y) & \cos(x'z)  \\
 \cos(\angle y'x) & \cos(\angle y'y) & \cos(y'z)  \\
 \cos(\angle z'x) & \cos(\angle z'y) & \cos(z'z)
\end{array}\right]$ is the matrix of the directional cosines,
$REF~=~\left[
\begin{array}{ccc}
  \dot\rho_s(\omega, \dot \varepsilon, \sigma) \cdot e^ {-i \phi} & 0 & 0 \\
  0 &\dot\rho_p(\omega, \dot \varepsilon, \sigma) \cdot e^ {-i\phi} & 0 \\
  0 & 0 & 1
\end{array}\right]$ is the transform reflected signal matrix,
where 
$\dot \varepsilon = \varepsilon_r- i 60 \sigma \lambda $ is the
dielectric index, $\sigma$ is the conductivity of the underlying
surface, $\phi=\omega \frac {2 h}{c} \cos(\angle z'z)$ is the
phase shift between direct and reflect signals, $h$ is the
altitude of the phase center of the dipole.

Finally, the values ${\dot E}_{x,y,z}^{space}(\omega)$ must be set
to the laboratory frame of reference, the center of which
coincides with the phase center of the telescope, and the axes
$x', y', z'$ correspond to directions: phase center-south, phase
center-west and phase center-zenith. This is achieved by using the
so-called directional cosines matrix ($COS$) which determines the
projection of the transformed components ${\dot
E}_{x,y,z}^{space}(\omega)$ to the new axes.

Equation (9) takes into account that the redistribution of the
polarization intensities of the incident wave at the receiver site
is not only due to the presence of different reflection
coefficients, but also due to different path lengths of the
incident and reflected waves. From equations (8) and (9) we see
the connection between the electrical field vector values at the
boundaries of homogeneous layers.

The simplest decision is to compensate for the impact of the
dispersion delay as shown in the paper of Hankins (1971). For the
coherent compensation of the dispersion delay it is sufficient to
multiply the matrix which includes factors of the dispersion delay
by its inverse matrix. Since the matrix, that characterizes this
effect in the chosen coordinate basis, is diagonal with a unit
determinant, the inverse matrix is simply complex conjugated to
the original matrix. This approach allows us to remove the
dispersion delay by layers.

Similarly, by using the multiplication to the inverse matrix, the
effect of the rotation of the polarization plane is compensated.
There is also a possibility of partial or gradual compensation of
the Faraday effect.

\sectionb{4}{THE METHODS OF THE ANALYSIS}

In order to increase the accuracy of the results, we carried out a
numerical simulation of the basic processes involved in the
generation of the PRE, in the processes of propagation in the cold
weakly anisotropic plasma and in the processing of the received
signal. For evaluation of the polarization parameters of the
received signals, we used a traditional method which uses the
polarization tensor (Zheleznyakov 1997). This method requires a
relatively long averaging in comparison with the characteristic
time of the fluctuations of amplitudes of the field components
(Mishchenko et al. 2006). When applied to the pulsar radio
emission, the polarization observations are very important as this
allows to probe the propagation medium and, in particular, the
ionosphere. For the decameter range it is more important to
investigate the anomalously strong pulses detected in a number of
pulsars (Popov et al. 2006; Ul'yanov \& Zakharenko 2012).

Since the use of the WF receivers is connected with necessity of
storing and processing a huge amount of information, the
registration with a parallel spectrum analyzer is mostly used. In
this case, the recorded signals contain information on its power
spectral density and on the dynamic behavior of the signal
envelope.

Hereafter we give the equations which are linking the electric
field wave intensity vector with the polarization tensor (Eq. 10)
and the Stokes parameters $I, Q, U, V$ (Eq. 11) in a circular and
linear cases:
\begin{equation}
J = \left[ \begin{array}{cc}                                                    
  I_{xx} & I_{xy} \\
  I_{yx} & I_{yy}
\end{array}
\right ]= \frac {c}{4\pi} \left ( \begin{array}{cc}
  \langle E_x(t)\cdot (E_x^*(t)) \rangle & \langle E_x(t) \cdot E_y^*(t)\rangle \\
  \langle E_y(t)\cdot E_x^*(t) \rangle & \langle E_y(t) \cdot E_y^* (t)\rangle
\end{array}  \right ){ ~~~},
\end{equation}
\begin{equation}                                                                
\begin{array}{l}
  I=I_{rr} +I_{ll} \\
  Q=I_{rl} +I_{lr} \\
  U=-i(I_{rl} -I_{lr}) \\
  V=I_{rr} - I_{ll}
\end{array}~,~~
\begin{array}{l}
  I=I_{xx} +I_{yy} \\
  Q=I_{xx} - I_{yy} \\
  U=I_{xy} +I_{yx} \\
  V=i(I_{yx} -I_{xy})
\end{array}{ }.
\end{equation}

When these parameters are known, we can determine the polarization
degree and PA (Zheleznyakov 1997).

These equations can be used when the received signals are being
recorded with a WF receiver. In the case of registering with a
spectrum analyzer, we can also recover all the Stokes parameters
of the original signal. Then the Faraday effect will cause the
polarization ellipse described by the electric intensity vector of
the received signal, being projected on a linear vibrator with
different position angles for different frequencies. Thus, if in
the total band of the reception signal PA of the polarization
ellipse makes $N$ turns to $\pi$ radians, the observer will
register the same $N$ periods of radiation intensity modulation in
the channel A or B (see Figure 2).

From the above considerations it is evident that the period of the
Faraday intensity modulation by the frequency $\Delta f_F$.
corresponds to a phase shift in the position angle $\Delta\psi
(RM, f)$ of $\pi$ radians.
\begin{figure}[h] \vbox{
\centerline{\psfig{figure=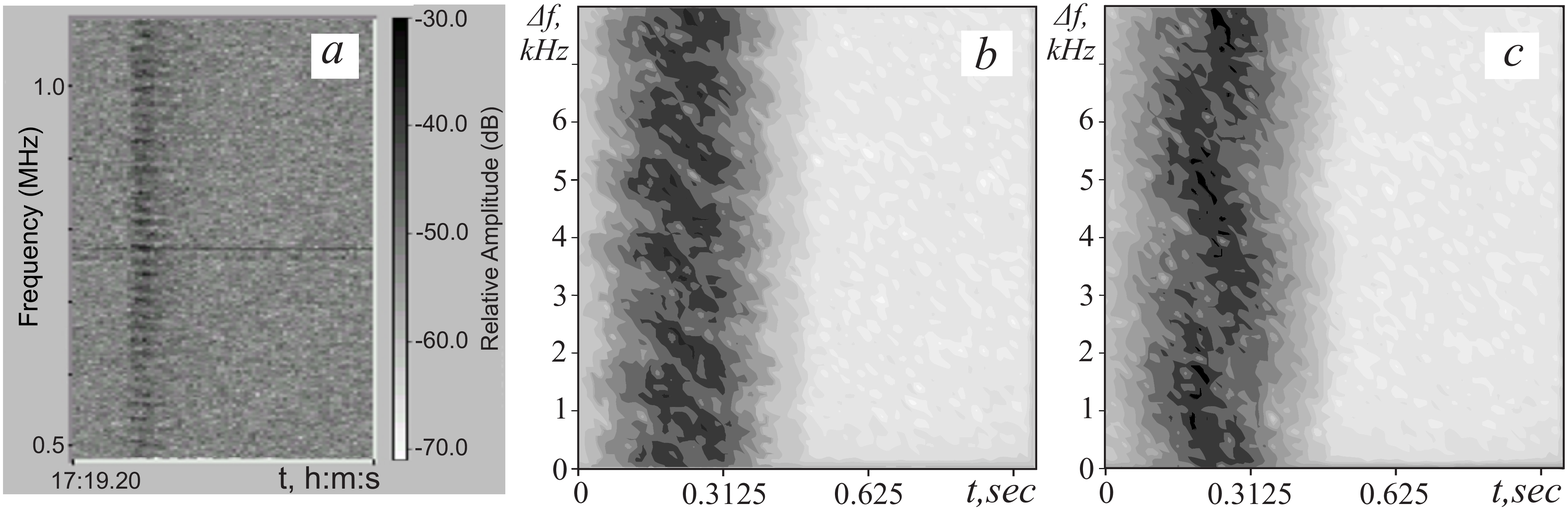,width=127mm,angle=0,clip=}}
\vspace{1mm} \captionb {2}{The dynamic spectra of:  a) the
elliptically polarized radiation from PSR B0809 +74,
 detected using UTR-2 ($F_c$  = 23.7 MHz,  $\Delta F$  = 1.53 MHz);
 b) the model signal ($F_c$= 20 MHz, $\Delta F$ = 4 kHz);
 c) the model signal ($F_c$= 30 MHz, $\Delta F$ = 4 kHz).}
 }
\end{figure}

\begin{equation}
\Delta \psi (RM, f)~=~c^2 RM \left[ \left( \frac{1}{f_c}\right) ^2-  
\left( \frac{1} {f_c+ \Delta f_F}\right) ^2 \right],~~ \Delta
\psi= \pi  {~~~}.
\end{equation}

The Faraday modulation intensity period is found from the number
$n_{max}$ of the spectral component for which the maximum power
response $(|\dot P(n_{max},t)|^2)$ is registered. Then the
estimate of $RM$ is calculated as:
\begin{equation}
RM=\frac{\pi}{c^2} \left [
 \frac{{f_c}^2 \left(f_c +\Delta f_F
\right)^2 } {\left(f_c +\Delta f_F \right)^2 -{f_c}^2}              
\right]~=~\frac{\pi}{c^2} \frac{{f_c}^3} {2 \Delta f_F}{~~~}.
\end{equation}

The errors of this method in percent are $\delta RM \leq 0.5 \cdot
100 /n_{max}$.
The phase of the registered spectral component will be linked to
PA of the polarization ellipse by the following relation:
\begin{equation}
\chi_{res}(t)=\frac{1}{2}\arg \left( \dot P (n_{max},t)\right){~~~},          
\end{equation}
where $\chi_{res}(t)$ is the PA of the restored signal.

\sectionb{5}{THE RESULTS}

The above methods were used to construct numerical models of the
polarized radio emission of pulsar pulses, as well as for the
treatment of three anomalously strong pulses of PSR B0809+74 which
were registered with a parallel spectrum analyzer. Figure 2 shows
the responses of the broad-band linear vibrators to the modeled
radiation pulses of pulsars, which passed through the interstellar
medium. These responses (Figure 2, panels b and c) are shown for
the receivers with the center frequencies 20 MHz and 30 MHz. The
dispersion measure for PSR B0809+74 ($DM=5.753 $~pc/cm$^3$) was
introduced into the ISM model, and then removed on the receiving
end according to the method described by Hankins (1971) and
Hankins \& Rickett (1975). Also, a deliberately high $RM = -234
$~rad/m$^2$ was introduced in the ISM model. This allowed us to
use the narrow-band polarization reception channels without any
loss of accuracy. In real observations, the bandwidth of the
receiving channels reaches several dozens of MHz what makes it
necessary to check the adequacy of the model assumptions in the
full bandwidth of the receiver.

Figure 3a shows the Stokes parameters of the model signal in the
reference frame connected with the pulsar (dash-dot line) and of
the same signal which has passed through the ISM with the given
and compensated $RM$ and $DM$ (solid line) parameters. Panel (b)
shows the trend of PA which is defined analytically (dashed line)
and was restored from the Stokes parameters, derived from the
polarization tensor in the solution of the inverse problem.
\begin{figure}[h] \vbox{
\centerline{\psfig{figure=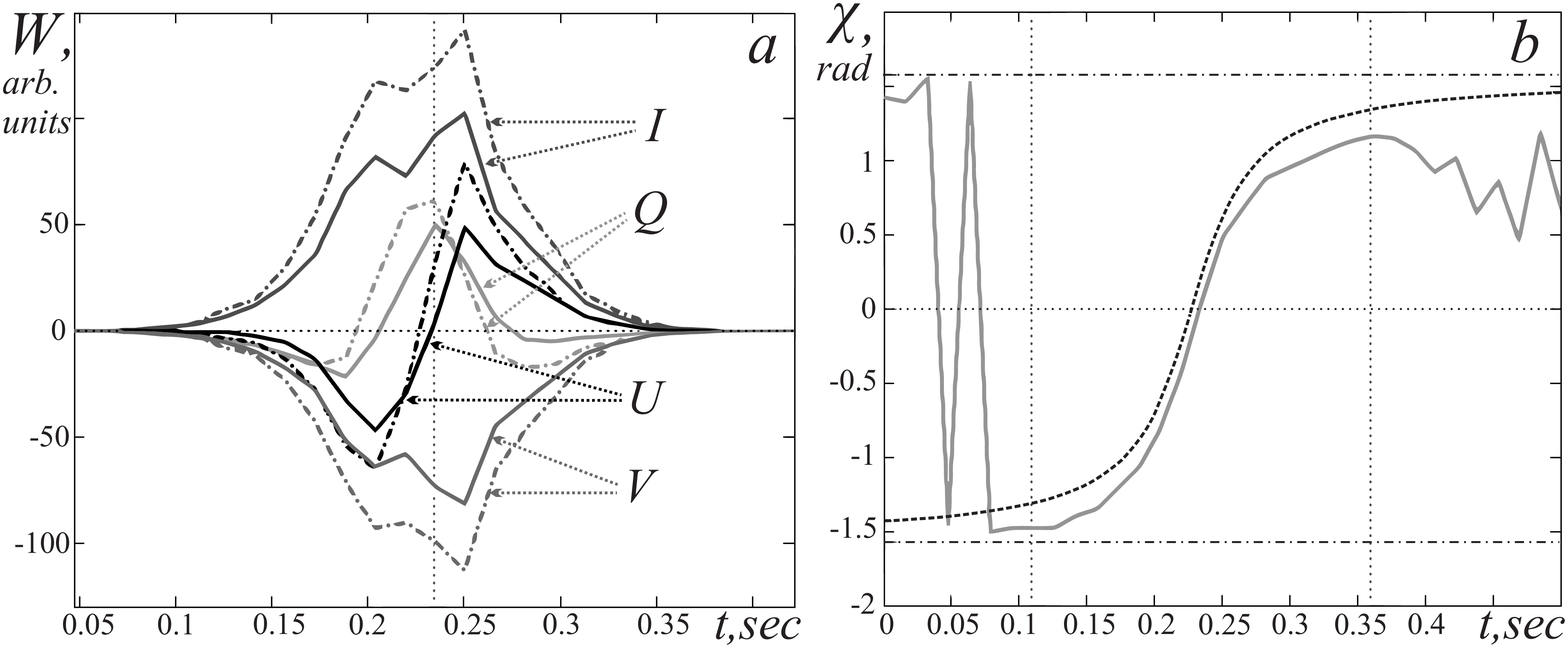,width=100mm,angle=0,clip=}}
\vspace{1mm} \captionb {3}{Panel (a) shows the Stokes parameters
and panel (b) - PA calculated from these parameters: $\chi =1/2
\arg(Q + iU)$.}
 }
\end{figure}

Three anomalously intensive pulses of PSR B0809+74 that have been
registered with the radio telescope UTR-2 and the broad-band
linear vibrators, have been processed using the methods described
above. In all three cases, the resulting estimate of $RM$ differs
from that which was obtained earlier from the high-frequency data,
and which we considered to be the aimed parameter of the ATNF
catalog (for PSR B0809+74 $~~RM_{ATNF}=-11.7 $~rad/m$^2$). Figure
4 shows the spectral response from which by Equation (13), we
obtain $RM$ estimates for each pulse.

The data on the obtained rotation measures and on their
determination errors are summarized in Table 1.
\begin{table}[h!]
\begin{center}
\vbox{\footnotesize\tabcolsep=3pt
\parbox[c]{124mm}{\baselineskip=10pt}
 {\smallbf\ \ Table 1.}{\small\
 The rotation measure of the individual pulses of the PSR B0809+74 at the frequency  $F_c  = 23.7 MHz$.\lstrut}}
\begin{tabular}{|p{0.060\linewidth}|p{0.060\linewidth}|p{0.100\linewidth}|p{0.110\linewidth}|p{0.100\linewidth}|p{0.080\linewidth}|p{0.100\linewidth}|p{0.100\linewidth}|}
\hline  {$Pulse$} & \centering { $n_{max}$} & \centering
{$RM_{Est}$, $rad/m^2$} & \centering {$RM_{ATNF}$, $rad/m^2$} &
\centering {$\Delta RM$, $rad/m^2$} & \centering {$\Delta RM$,
$\%$} & \centering {$\delta RM$, $rad/m^2$} & ~$\delta RM$,
$~~~~~~\%$ \\
 \hline
 \centering {1~(a)} & \centering {80} & \centering {-12.18} & {~} &\centering {-0,48} & \centering {4.1} & \centering {±0.073} &
 ~~{0.625}
 \\ \cline{1-3} \cline{5-8} \centering {2~(b)} & \centering
{83} &\centering {-12.64} & \centering {-11.7} &\centering {-0.94}
&\centering {8} &\centering {±0.075} &
 ~~{0.602}
\\ \cline{1-3} \cline{5-8} \centering {3~(c)} &\centering {83} &\centering {-12.64} & {~} &\centering {-0.94} &\centering {8} &\centering
{±0.075} & ~~{0.602} \\ \hline
\end{tabular}
\end{center}
\end{table}

In Table 1 the following designations are used: $\Delta
RM_{rad/m^2} = RM_{Est}- RM_{ATNF}$; $\delta RM_{\%} = 0.5 \cdot
100/ n_{max}$; $\delta RM_{rad/m^2} = \delta RM_{\%} \cdot
RM_{ATNF}/100$.
\begin{figure}[h!] \vbox{
\centerline{\psfig{figure=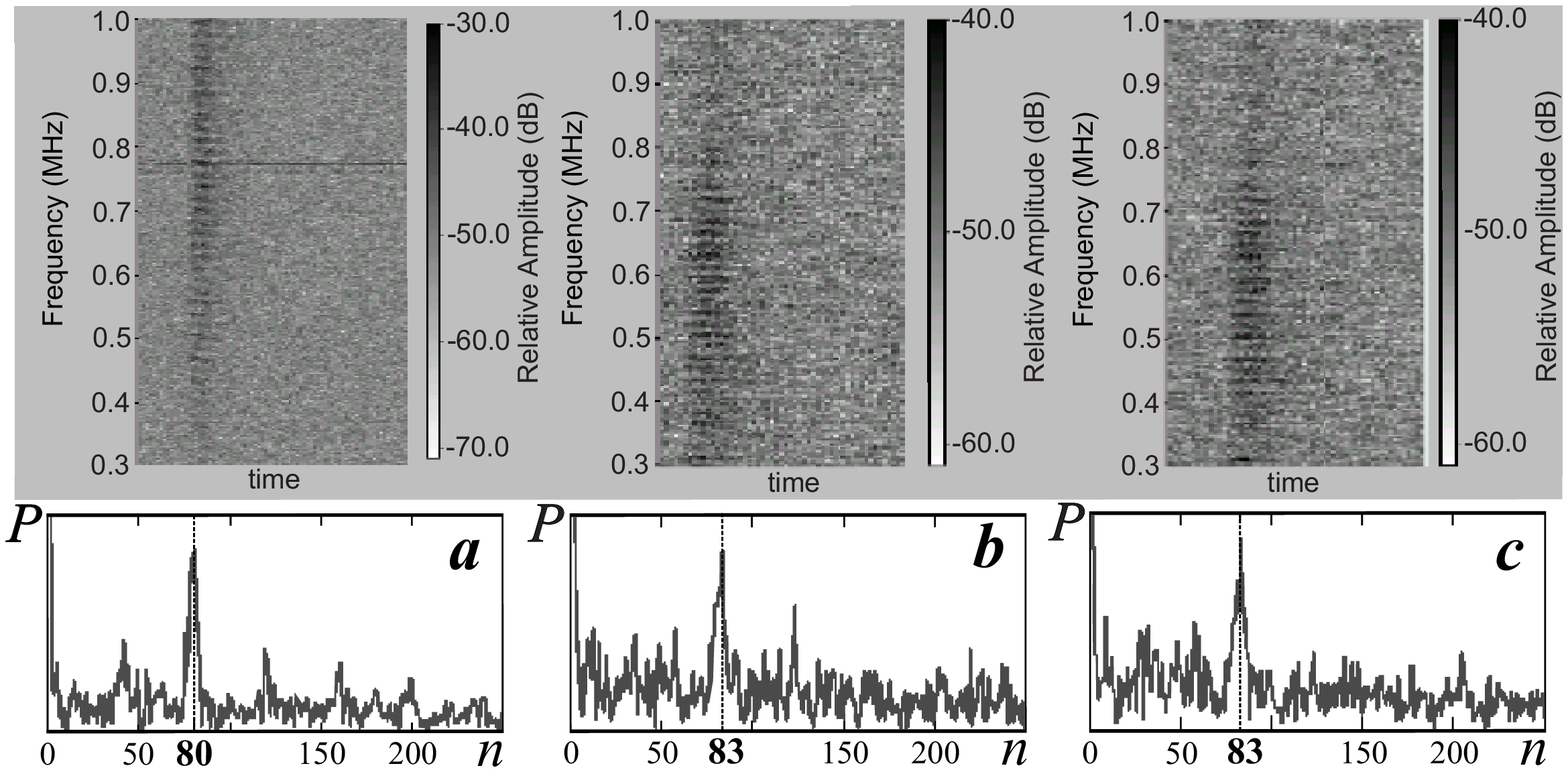,width=127mm,angle=0,clip=}}
\vspace{1mm} \captionb {4}{ Top panels. Dynamic spectra of the
elliptically polarized radiation of the three anomalously strong
pulses of PSR B0809+74 registered with UTR-2 ($F_c$  = 23.7 MHz,
$\Delta F = 1.53$ MHz). Lower panels.
 The polarization responses ($P~=~|\dot P(n,t)|^2$) near the maximum intensity of each pulse.}
}
 \end{figure}

From the results obtained from the variations of the rotation
measures for the three individual pulses we conclude that the
accuracy of their registering does not cause any doubt, since the
errors of the method are smaller by an order of magnitude than the
variations of $RM$. Since these variations are registered along
the whole way of the radiation from the pulsar to the observer, we
should be in caution in their interpretation. In this case, in the
reference frame of the observer, the variations were fixed at the
time of the dispersion delay of each individual pulse. For the
specific observations, this time delay was equal to 5 s. The
responsibility for such variations of the $RM$ can carry the
variations of the magnetic field and the electron density in the
magnetosphere of the pulsar itself, as well as the variations of
the same quantities in the Earth ionosphere. In order to
distinguish the place of these variations along the line of sight,
it is appropriate to lead simultaneous broad-band observations of
pulsars, including the high-frequency bands. On the other hand, a
comparison of the given data with the two-frequency data obtained
from the analysis of the difference between the time of pulse
arrival of the GPS system and the same signals of the local
hydrogen standards, are needed.

High-frequency data obtained for pulsars, allows us to
significantly reduce the time of the dispersion delay, while the
two-frequency data comparing the GPS time scales and the hydrogen
standards will help us to take into account the variations of
electron density in the line of sight in the ionosphere. In
general, this will allow us to specify the area of the rapid
magnetic field fluctuations along the line of sight. Inclusion in
the analysis of a system of many pulsars simultaneously increases
the number of directions on which we can study the propagation
medium, including the Earth ionosphere. Ultimately, this approach
will allow passing to the construction of models of the ionosphere
with dynamically varying parameters. This may be resulted in the
further development of methods of adaptive radio astronomy
described by Afraimovich (1981, 2007).


\sectionb{6}{CONCLUSIONS}

The layered model of the propagation medium, which can be used
together with the eikonal equation, allows us to split the time
delay and the angles of rotation of the polarization plane.
Formally, this leads to the appearance of the equations of the
commuting matrices, one of which describes the effect of the delay
dispersion in a cold plasma, and another one describes the Faraday
effect.

Systematic observations of radio pulses of pulsars in the mode of
waveform recording or the spectra analyzer mode, together with the
data of the dualfrequency GPS monitoring, can more accurately
determine the total electron contents in the column of the
ionosphere along the line of sight, including the rapid
fluctuations of this parameter. Analysis of the polarization
parameters derived from these observations allows us to estimate
the slow and rapid changes in the magnetic field vector and to
include this information to the dynamical model of the ionosphere.

\thanks{ The authors are thankful to Zakharenko V.V. \& Deshpande A.A.  for their help in the carrying outing of the observations. The use of
the ATNF databases is acknowledged.}

\References

\refb Afraimovich E. L., 1981, Astronomy \& Astrophysics, 97, 366

\refb Afraimovich E. L., 2007, Rep. of Acad. Science, Moscow, 417,
818 (in Russian)

\refb ATNF catalog http://www.atnf.csiro.au/people/pulsar/psrcat/

\refb Eisenberg G. Z, Belousov S.P., Zhurbenko E.M. et al., 1985,
\textit {Short-wave antennas}, Radio and Communication Publishers,
Moscow, 2nd edition (in Russian)


\refb Ginzburg V. L., 1967, \textit {Propagation of
Electromagnetic Waves in Plasmas}, Nauka Publishers, Moscow

\refb Hankins T. H., 1971, ApJ, 169, 487

\refb Hankins T. H., Rickett B. J., 1975, in \textit {Methods in
Computational Physics}, vol. 14, Radio astronomy, Academic Press,
p. 55

\refb Hewish A., Bell S. J., Pilkington J. D. H. et al., 1968,
Nature, 217, 709

\refb Karastergiou A. and Johnston S. 2006, MNRAS, 365, 353

\refb Marple S.L. Jr., 1999, IEEE Transactions on Signal
Processing, 47, 2600

\refb McKinnon M.M., 2009, ApJ, 692, 459

\refb Melrose D., 2003, in \textit {Radio Pulsars}, ASPC, 302, 179

\refb Melrose D., Luo Q., 2008, in \textit {40 Years of Pulsars:
Millisecond Pulsars, Magnetars and More}, AIPC, 983, 47

\refb Mishchenko, M.I., Travis, L.D., Lacis, A.A., 2006, \textit
{Multiple Scattering of Light by Particles}, Cambridge University
Press

\refb Petrova S.A., 2006a, MNRAS, 366, 1539

\refb Petrova S.A., 2006b, MNRAS, 368, 1764

\refb Pilkington J.D.H., Hewish A., Bell S.J., Cole T.W., 1968,
Nature, 218, 126

\refb Popov M.V., Kuz'min A.D., Ul'yanov O.M. et al., 2006, Astr.
Rep., 50, 562

\refb Schelkunoff S.A., Friis H. T., 1952, \textit{Antennas:
Theory and Practice}, Bell Telephone Laboratories, New York : John
Wiley \& Sons

\refb Suleimanova S.A., Pugachev V.D., 2002,  Astr. Rep., 46, 309

\refb Ulyanov O.M., 1990, Kinematics and Physics of Celestial
Bodies, Kiev, 6, 63

\refb Ul'yanov O.M., Zakharenko V.V., Konovalenko A.A. et al.,
2006, Radio Physics and Radio Astronomy (in Russian), 11, 113

\refb Ul'yanov O.M., Zakharenko V.V., Bruk Yu.M., 2008, Astr.
Rep., 52, 917

\refb Ul'yanov O.M., Zakharenko V.V., 2012, Astr. Rep., 56, 417

\refb Wang C., Han J., Lai D., 2011, MNRAS, 417, 1183

\refb Zakharenko V.V., Nikolaenko V.S., Ulyanov O.M. et al., 2007,
Radio Physics and Radio Astronomy (in Russian), 12, 233

\refb Zheleznyakov V.V., 1977, \textit{Electromagnetic Waves in
Cosmic Plasma. Generation and Propagation }, Nauka Publishers,
Moscow (in Russian)

\refb Zheleznyakov V. V. 1996, \textit{Radiation in Astrophysical
Plasmas}, Astrophysics and Space Science Library, vol. 204, Kluwer
Academic Publishers, Dordrecht (published in Russian in 1997)

\end{document}